\documentstyle[prb,twocolumn,aps,graphicx]{revtex}
\begin{document}
 \newcommand{\mytitle}[1]{
 \twocolumn[\hsize\textwidth\columnwidth\hsize
 \csname@twocolumnfalse\endcsname #1 \vspace{1mm}]}
\mytitle{
\title{Interference and interaction effects in multilevel quantum dots}
\author{Daniel Boese,$^{1,2}$ Walter Hofstetter,$^{3}$ and 
Herbert Schoeller$^{2,1}$\cite{new1}}
\address{$^1$ Institut f\"ur Theoretische Festk\"orperphysik, Universit\"at
  Karlsruhe, D-76128 Karlsruhe, Germany\\
$^2$ Forschungszentrum Karlsruhe, Institut f\"ur Nanotechnologie, D-76021
Karlsruhe, Germany\\
$^3$ Theoretische Physik III, Elektronische Korrelationen und Magnetismus, 
Institut f\"ur Physik, Universit\"at Augsburg, D-86135 Augsburg, Germany}
\date{\today}
\maketitle
\begin{abstract}
Using renormalization group techniques, we study spectral 
and transport properties of a spinless interacting quantum dot consisting
of two levels coupled to metallic reservoirs. For strong Coulomb
repulsion $U$ and an applied Aharonov-Bohm phase $\phi$, we find a 
large direct tunnel splitting 
$|\Delta|\sim (\Gamma/\pi)|\cos(\phi/2)|\ln(U/\omega_c)$ between the
levels of the order of the level broadening $\Gamma$.
As a consequence we discover a many-body resonance in the spectral density
that can be measured via the absorption power.
Furthermore, for $\phi=\pi$, we show that the system can 
be tuned into an effective Anderson model with spin-dependent tunneling.
\end{abstract}
}
{\em Introduction.}
Electronic transport through ultra-small quantum dots (QD), where the charging
energy is the largest energy scale, has been studied extensively over 
the last few years.\cite{curacao} Due to the quantization of charge the
transport is dominated by Coulomb blockade (CB). More recently experiments
revealed that the transport can be even more intriguing by measuring the Kondo
effect \cite{kondo-exp,simmel} as suggested in Ref.'s.\cite{kondo-theo}

The Kondo effect occurs for a dot with one low-lying
spin-degenerate level. In this paper, we will study 
a dot consisting of two levels without spin or, 
equivalently, two dots in an Aharonov-Bohm (AB) geometry with one
level per dot in the presence of an interdot Coulomb repulsion $U$. 
Such a system is of fundamental interest since the
two possible paths through the dot (via level $1$ or
$2$) can interfere with each other. The interference can be controlled
by an AB flux and has attracted much interest due to
the possibility of realizing AB interferometers \cite{yacoby}
or use the coherent properties in connection with quantum
computing \cite{loss} (for recent experimental realizations see
Ref.~\cite{holleitner-etal}). Furthermore, in many recent
experiments performed in the strong tunneling regime, the
level broadening is large and transport is inevitably 
controlled by multilevel physics.

{\em The model.}
We consider a quantum dot consisting of two levels, labeled 
by $j=1,2$. Via tunnel barriers the dot is connected 
to two electronic reservoirs $r=L,R$. The orbital index $j$ is not
conserved during tunneling and hence does not exist in the
leads. The Hamiltonian is written as 
$H = H_{\mathrm dot} + H_{\mathrm res} + H_T$, with
$H_{\mathrm dot} = \sum_j \varepsilon_j c^\dagger_j c_j + U n_1 n_2$, 
$H_{\mathrm res}=\sum_{kr}  \varepsilon_{kr} a^\dagger_{kr} a_{kr}$, and
$H_T = \sum_{rkj} (t^r_j a^\dagger_{kr} c_j + {\mathrm H.c.})$.
The tunnel matrix elements are assumed to be real except for an
AB phase, i.e., we attach a phase factor $e^{i\phi}$ to
$t_2^L$. The energy scale of
the level broadening is defined by $\Gamma_j^r = 2\pi |t_j^r|^2\rho_0$, 
where $\rho_0$ is the density of states in the leads, which we assume to be
independent of energy for the energy range of interest. 

We neglect spin (assuming a large Zeeman splitting)
since the aim is at analyzing explicitly the physical effects arising from 
the tunneling--induced interference between the two levels. 
Since both levels overlap
with the reservoir states, there is an effective overlap matrix
element $-\Delta/2$, which, surprisingly,
is shown to be zero for a noninteracting quantum dot, but for strong 
on-site Coulomb repulsion $U>>|\epsilon|,\Gamma$, scales like
\begin{equation}\label{Delta}
\Delta\sim {\sqrt{\Gamma_1^R\Gamma_2^R}+
            \sqrt{\Gamma_1^L\Gamma_2^L}e^{i\phi}\over \pi}
            \ln{(U/\omega_c)}\,.
\end{equation}
Here, $\phi$ is the AB phase, and $\omega_c$ denotes a low-energy cutoff 
set by the maximum of the mean level position 
$\epsilon=(\epsilon_1+\epsilon_2)/2$, the mean level broadening 
$\Gamma=(\Gamma_1+\Gamma_2)/2$ (with $\Gamma_j=\Gamma_j^R+\Gamma_j^L$), 
the temperature $T$, or the bias voltage $eV$. The level splitting is 
given by
\begin{equation}\label{splitting} 
\delta\tilde{\epsilon}=\sqrt{\delta\epsilon^2 + |\Delta|^2}\,, 
\end{equation}
where $\delta\epsilon=\epsilon_2-\epsilon_1$ denotes the level spacing. 
Consequently, the tunnel splitting gives rise to an 
interference-- {\it and} interaction--induced level repulsion, i.e., an 
effect not being considered in models with levels labeled by a 
conserved quantum number (e.g.~spin) \cite{kondo-theo} or in the absence of 
interactions.\cite{koenig-gefen} The energy scale of $\Delta$ 
is given by $\Gamma$ and will influence the spectral 
properties as well as the conductance for low enough temperatures
$T\lesssim\Gamma$. We 
emphasize that this energy scale is well separated from
the Kondo temperature $T_K\sim \sqrt{\Gamma U}\exp(\pi\epsilon/\Gamma)$
($\epsilon\ll -\Gamma$), which is exponentially small and determines the
crossover to the occurrence of the Kondo effect for spin-degenerate
levels \cite{kondo-theo}. Most importantly, we will show in this paper that
for low lying levels $\epsilon\lesssim -\Gamma$ (where the ground state 
is the singly occupied state),
the effective level splitting shows up in a many-body resonance in
the spectral density at the energy $\delta\tilde{\epsilon}$, which 
e.g.~can be measured by an absorption experiment but influences
also the temperature and flux dependence of the linear conductance.
For $\phi=\pi$ and $\Gamma^R_j=\Gamma^L_j$, the tunnel splitting is 
zero, and the system is shown to be equivalent to an Anderson model 
with Zeeman splitting $\delta\epsilon$. Thus, Kondo physics can be 
realized in a quantum dot without spin even if the 
quantum number labeling the levels is {\it not} conserved.

We note that multilevel dots in the presence of spin have been 
studied previously.\cite{pohjola,inoshita,izumida1} However,
Ref.~\onlinecite{pohjola} studies the case of a conserved quantum 
number labeling the levels, and Refs.~\onlinecite{inoshita} and
\onlinecite{izumida1}  
consider the cases $\delta{\epsilon}\gg\Gamma$ or $\phi=\pi$, where 
the effect of the tunnel splitting $\Delta$ can be neglected.
The same applies to the AB geometry of Ref.~\onlinecite{izumida2}
where the interdot Coulomb repulsion is absent.

{\em Renormalization group study.}
An effective dot Hamiltonian can easily be derived from perturbation
theory or, equivalently, by integrating out the reservoir states
by the renormalization group. The dot is characterized
by four states $|0\rangle$, $|1\rangle$, $|2\rangle$, and $|12\rangle$, 
with energies $E_0=0$, 
$E_1=\epsilon_1$, $E_2=\epsilon_2$, and $E_{12}=\epsilon_1+\epsilon_2+U$. 
Intuitively, the hybridization with the reservoirs will lower the
energies of all these states. For the singly occupied states, however, this is
less  pronounced because it costs a finite energy $U$ to occupy the dot with
a second electron. Therefore, the level positions $\epsilon_j=E_j-E_0$
will be renormalized upwards. Furthermore, a coupling between the
levels is generated since tunneling events can shift the electrons
between the two levels. For an electron starting in level 1 there are
two possibilities: either the electron first tunnels out and hops into level
2 or an electron first hops into level 2 and then the electron tunnels
out of level 1. In the latter case, the intermediate state is the doubly
occupied state and, due to Fermi statistics, the matrix element gets
an additional minus sign. Therefore, for reservoir electrons with
an energy $|\epsilon_k|>>U$, these two terms will cancel each other
and, consequently, there is no direct coupling between the levels in
the noninteracting case. In contrast, for an interacting system,
the doubly occupied state is suppressed, and there is a finite coupling
$\Delta$ between the two levels. We note that this mechanism does
not work for levels characterized by spin since the two tunneling
processes described above would also change the spin in the reservoirs
and, therefore, do not lead to a direct renormalization of the
dot Hamiltonian. 

Using the real-time renormalization group (RG) of Ref.~\onlinecite{schoeller-RG} 
for the forward propagator we find that energy scales 
$\omega_c > U$ do not renormalize the states, i.e., we start the RG 
at $\omega_c={\rm min}(D,U)$ where $D$ is the bandwidth.  
In the basis of the three remaining states $|0\rangle$,
$|1\rangle$, and $|2\rangle$, we obtain the flow equation ($t_c=1/\omega_c$)
\begin{equation}\label{RG}
\frac{dH_{\mathrm dot}}{dt_c}
= -{1\over 2\pi (t_c-i0^+)}
\left(\begin{array}{ccc}
\Gamma_1 + \Gamma_2 & 0 &0 \\
0 & \Gamma_1 & \Phi \\
0 & \Phi^\ast & \Gamma_2
\end{array}\right) \; ,
\end{equation}
where $\Phi=\sqrt{\Gamma_1^R\Gamma_2^R}+\sqrt{\Gamma_1^L\Gamma_2^L}e^{i\phi}$. 
Neglecting level broadening, the solution of this equation
gives an upward level shift $E_{1/2}-E_0=\epsilon_{1/2}+\lambda\Gamma_{2/1}$, 
with $\lambda=(1/2\pi)\ln(U/\omega_c)$, and a coupling 
$-\Delta/2=-\Phi\lambda$ leading to Eq.~(\ref{Delta}). 
As a consequence we get two effective levels at 
$\tilde{\epsilon}_{1/2}=\epsilon+\lambda\Gamma\mp\delta\tilde{\epsilon}/2$,
where the effective level splitting $\delta\tilde{\epsilon}$ is given by
Eq.~(\ref{splitting}). While $\tilde{\epsilon}_1$ is quite close to the
original level position, $\tilde{\epsilon}_2$ is strongly renormalized 
upwards. For $\Gamma_j^r\approx\Gamma/2$, $\delta \epsilon \ll \Gamma$ and 
$\phi\ll 1$ the lower (upper) level is coupled strongly (weakly) to the 
reservoirs.
For the following discussion we will usually assume equal couplings
$\Gamma_j^r=\Gamma/2$, i.e., $|t_j^r|=t_j=t$ and discuss the effect of 
asymmetries at the appropriate places. 

In the symmetric case we define $\sqrt{2}f_i=c_1-(-1)^i c_2$.
For $\phi=0$, only the
$f_1$ operator couples to the reservoirs, whereas the level spacing
$\delta\epsilon$ controls the coupling between the $f_1$ and $f_2$
level. The current operator in the right reservoir is given by
$I_R=ie\sqrt{2}t\sum_k (a^\dagger_{kR}f_1-{\mathrm H.c.})$.
We also note that for $\delta\epsilon=0$ and $\phi=\pi$ 
the conductance is exactly zero since the $f_1$ ($f_2$) level couples 
only to the left (right) reservoir. This is an effect of destructive
interference which interestingly persists also in the presence of
interactions.

{\em Spectral density and absorption power.}
In Fig.~\ref{figure1} we show the
spectral density of the $f_1$ level for $\epsilon>0$, where the 
ground state is given by the empty state.
The results are obtained by using the full real-time renormalization
group method of Ref.~\onlinecite{schoeller-RG} which is known to yield
excellent results in the regime where charge fluctuations dominate.
The two peaks in the spectral density correspond to the
renormalized level positions and change qualitatively as function
of temperature $T$ and $\delta\epsilon$ according to 
Eqs.~(\ref{Delta}) and (\ref{splitting}). 
The distance between the resonances saturates for 
$\delta{\epsilon}<\Gamma$ at the energy scale $\Delta$ according
to Eq.~(\ref{splitting}). In contrast, when $\epsilon$ is below the 
Fermi level, the lower level is occupied and particle excitations lead to a 
broad shoulder in the spectral density at the effective
spacing $\delta\tilde{\epsilon}$, see inset of Fig.~\ref{figure2}
(an additional weak feature occurs at negative frequencies
but this is masked by the broad resonance at $\tilde{\epsilon}_1$). 
These results have been obtained by using Wilson's nonperturbative 
numerical renormalization group (NRG) \cite{NRG} which, 
up to some overbroadening effects at higher  
frequencies, gives very precise results for the spectral density
near the Fermi level and for the positions of all resonances. 
Since the location of the shoulder is not
at the Fermi level, it is more suitable to test its position via
the absorption power rather than the linear conductance.
Therefore, we have shown in Fig.~\ref{figure2} the result
for the spectral density of the transition operator 
$c_1^\dagger c_2 + c_2^\dagger c_1$. The peak position of the
absorption power agrees very precisely with the position 
of the shoulder in the spectral density of the $f_1$ level.
We emphasize that the shoulder is absent without the tunnel
splitting, i.e., it is a generic effect which will also be 
present in the asymmetric case $\Gamma_1\ne\Gamma_2$. In this
case, however, the broadening of the shoulder (which is determined 
by $\Gamma$) will increase relative to its height (which is 
determined by $\Delta$).

Figure \ref{figure3} shows the spectral density of the $f_1$ level for
different AB phases $\phi$ (also obtained by NRG). For $\delta\epsilon=0$, 
the position of the shoulder varies proportional to 
$|\Delta|\sim (\Gamma/\pi)|\cos(\phi/2)|\ln(U/|\epsilon|)$,
according to Eq.~(\ref{Delta}) (for $\phi=0$, the amplitude of the shoulder
is zero since the $f_2$ level is decoupled from the reservoirs).
Furthermore, the resonances at finite frequency 
become more pronounced and, for $\phi=\pi$, 
merge into a Kondo resonance at the Fermi level
with width given by the Kondo temperature $T_K$. This effect can
easily be understood, since for $\phi=\pi$, the tunneling Hamiltonian 
reads $H_T=t\sum_{kj}(b_{kj}^\dagger c_j + {\mathrm H.c.})$ with
$\sqrt{2}\,b_{ki}=a_{kR}-(-1)^i a_{kL}$.
Hence, for this special case, the pseudo spin $j$ is 
effectively a conserved quantum number and we obtain the Hamiltonian of the
usual Anderson model, which, for a low lying level
$\epsilon$ is equivalent to the Kondo model.\cite{hewson} 

We note that this realization of Kondo physics without explicit spin
degrees of freedom is quite different from other realizations, 
where metallic\cite{matveev} or two-level systems\cite{zawadowski} 
have been used. Furthermore, there are three experimentally 
tunable ways to destroy the Kondo resonance.
First, a finite level spacing $\delta\epsilon\ne 0$ acts like an 
effective Zeeman splitting. This splits the Kondo resonance and 
decreases its height, see the left inset of Fig.~\ref{figure3}. 
Second, an AB phase away from $\phi=\pi$ leads to an effective coupling 
$\Delta$ between the two levels. At $\Delta\sim T_K$ a phase
transition will occur quite analog to the competition between
RKKY and Kondo physics in two-impurity models.\cite{hewson,ruderman}
The same mechanism is induced by left/right asymmetries,
i.e., for $\Gamma_j^R\ne\Gamma_j^L$. Third, for given left/right 
symmetry but $\Gamma_1\ne\Gamma_2$, we obtain an Anderson model
with pseudo-spin-dependent tunneling matrix elements $t_j$.
As shown in the right inset of Fig.~\ref{figure3}, the Kondo resonance 
arising at $\Phi = \pi$ is reduced and splits asymmetrically 
but is well defined even at $\Gamma_2/\Gamma_1 \approx 2$ (we note 
that the reduction is quite more pronounced for a finite level spacing). 
As a consequence, the Kondo resonance can be shifted away from the 
Fermi level by changing the asymmetry of the tunneling matrix elements,
an effect also seen in recent experiments.\cite{simmel}

Since the conductance is zero for $\delta\epsilon=0$ and $\phi=\pi$, the 
Kondo resonance will show up only weakly in the $I(V)$ characteristics 
by changing the level spacing or the AB flux. However, the crossover
to the Kondo effect can e.g.~be measured by the absorption power.
Alternatively, in an AB geometry with two dots and one level per dot 
we expect in equilibrium for $\delta\epsilon=0$ and $\phi=\pi$ a Kondo 
resonance in each dot separately. Their effect might be tested
by measuring the conductance fluctuations of a parallel quantum wire 
lying very close to one dot.

{\em Linear conductance.}
Another fingerprint for the renormalization of the energy levels
due to the tunneling splitting $\Delta$ is the measurement of the
linear conductance. It is calculated by using the renormalized 
Hamiltonian on the
forward and backward propagator according to Eq.~(\ref{RG}), 
including the level broadening (for the backward propagator we take 
the hermitian conjugate\cite{schoeller-RG}). This effective
Hamiltonian is used as an input for the calculation of rates
in lowest order in the tunneling coupling. 

Figure \ref{figure4} shows the temperature dependence for 
$\delta\epsilon=\phi=0$.\cite{exsolv}
At $T=0$, the spectral density of the $f_1$ level is a single Lorentzian 
with width $\Gamma$
centered at the level position $\epsilon$. The resonance at $\epsilon+U$ is 
missing for $\epsilon>-U/2$ since the $f_2$ level is decoupled from the 
system and is not occupied in the ground state; see also inset of 
Fig.~\ref{figure2}. Thus, at zero temperature, 
the conductance is symmetric under a sign change of $\epsilon$, in contrast
to the case for spin degenerate levels, where the Kondo effect enhances
the conductance for negative $\epsilon$. For finite temperature, the
$f_2$ level starts to become occupied and suppresses the conductance
due to the Coulomb repulsion $U$. This effect is more pronounced for
negative $\epsilon$ and, therefore, the conductance shows a local maximum
for $T\sim\epsilon>0$ but is nearly monotonic for 
$\epsilon<0$. This distinguishes the model from transport through a
single level.

The inset of Fig.~\ref{figure4} shows the gate voltage dependence for different
AB phases and $\delta\epsilon=0$. 
The RG predicts $\Delta$ to decrease with increasing flux.
For $\phi\rightarrow\pi$ the tunnel splitting is small and the level 
shift by $\lambda$ leads to a resonance position of the linear conductance 
near $\epsilon=-\lambda$. In contrast, for $\phi\rightarrow 0$, the
tunnel splitting is large and $\tilde{\epsilon}_1\approx\epsilon$ which
leads to a resonance position near $\epsilon=0$. As a consequence we find 
that the position of the resonance is strongly influenced by the AB phase
and reflects directly the tunnel splitting $|\Delta|$ together with the
level renormalization $\lambda$.

Finally, we would like to comment on the case when the number of levels
is given by $N>2$. Generalizing the RG equation (\ref{RG}) to this case 
gives rise to an upward shift of all particle and hole excitations by
approximately $\sim N\Gamma$, while only one level with an equally increased 
broadening remains approximately at the original position. This means that 
transport appears to be effectively controlled by single-level physics
and may explain recent experiments\cite{kondo-exp,simmel} in the regime 
$\Gamma\sim\delta\epsilon$ where universal Kondo behavior of single-level
dots has been observed. 

{\em Summary.} 
We have studied interaction and interference effects in quantum dots
with two levels or two quantum dots with one level coupled to reservoirs. 
We found a new tunnel splitting that changes as a function of an applied 
magnetic flux and can be measured via the absorption power. As function 
of the flux, the system can be tuned into 
an effective model showing Kondo physics. We expect important implications 
of our results for transport and spectroscopy experiments as well as for
the theory of level statistics in quantum dots.

We would like to thank J\"urgen K\"onig, Teemu Pohjola, and Gerd Sch\"on
for valuable discussions. This work is supported by the
DFG as part of the Graduiertenkolleg ''Kollektive Ph\"anomene im
Festk\"orper'' (D.B.), "SFB 195" (D.B. and H.S.) and "SFB 484" (W.H.).

\begin{figure}
  \centerline{\includegraphics[width=7cm]{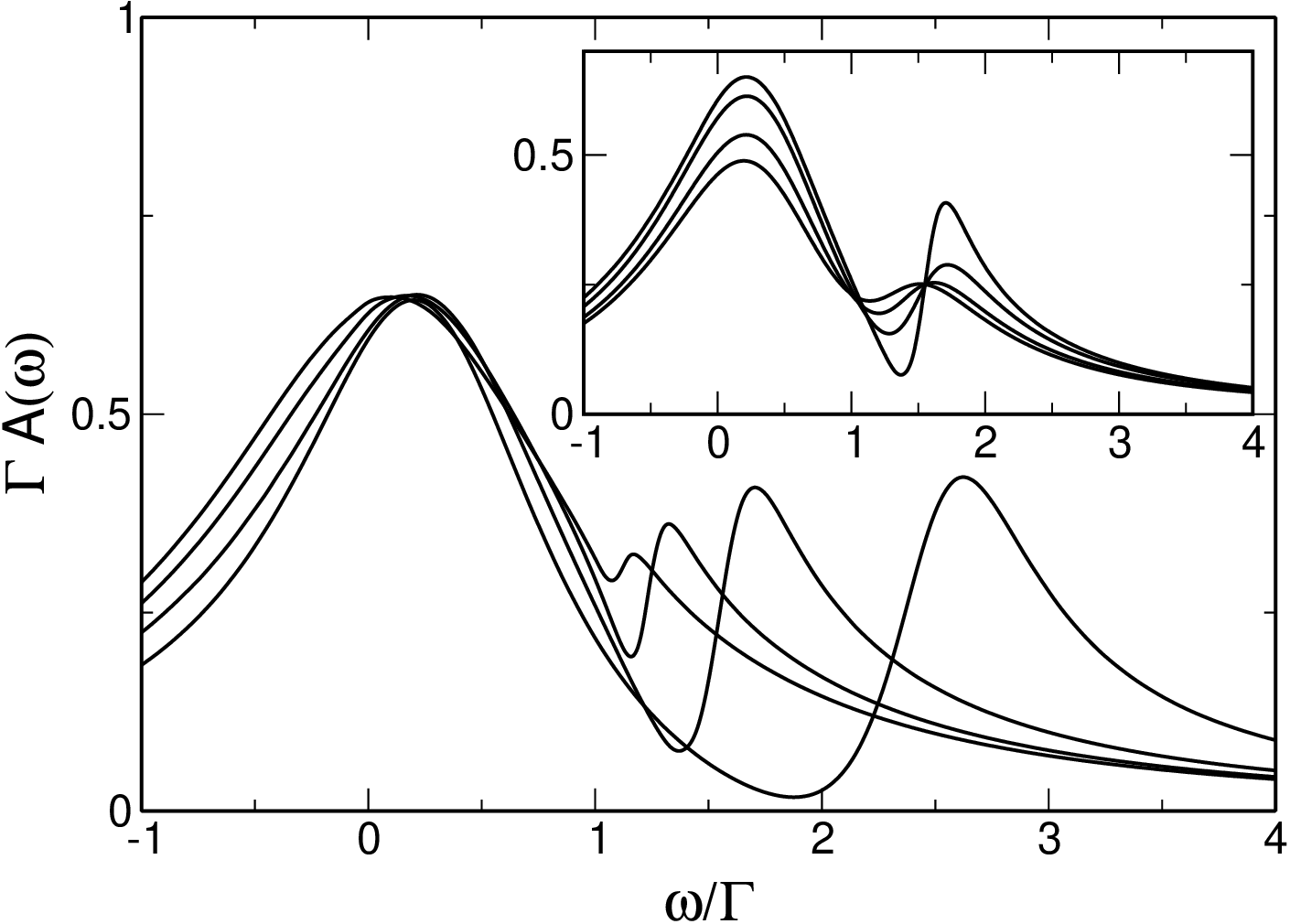}}
        \caption{Spectral density of the $f_1$ level for 
          $\epsilon_1=T=0$, $U=10\Gamma$, 
          $\delta\epsilon=0.25,0.5,1,2\Gamma$ (from left to right). 
          Inset: $\delta\epsilon=\Gamma$, $T=0,0.5,1,1.5\Gamma$ 
          (from top to bottom).}
        \label{figure1}
\end{figure}
\begin{figure}
  \centerline{\includegraphics[width=7cm]{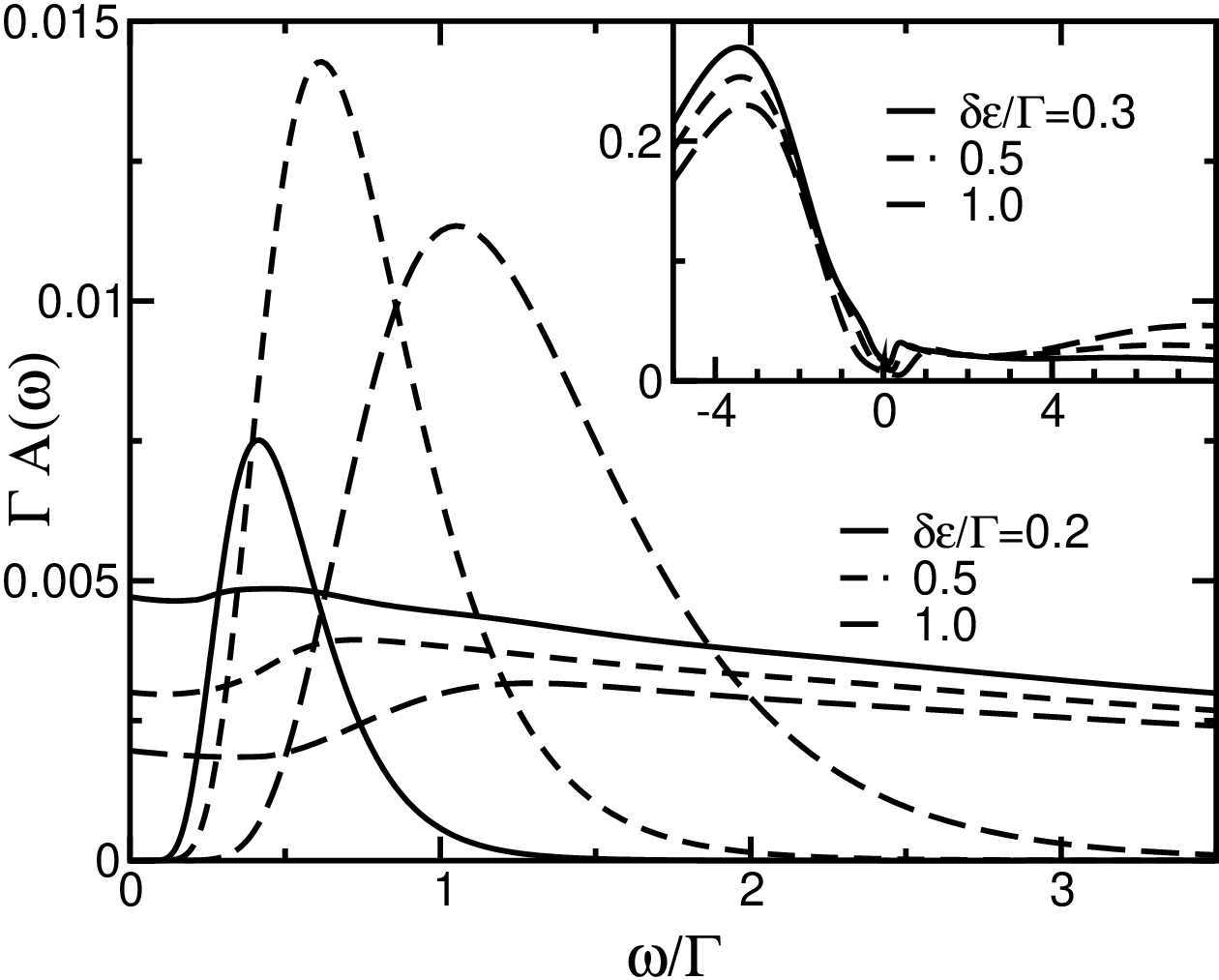}}
        \caption{Absorption power (scaled, single peaks) 
vs.~spectral density of the $f_1$ level (``shoulder'' at the same position) 
for $\epsilon_1 = -10 \Gamma$ , $U=50 \Gamma$, and $T=0$. 
In the inset, the spectral density is shown 
for $\epsilon_1 = -3.5 \Gamma$ and $U=10 \Gamma$.}
        \label{figure2}
\end{figure}
\begin{figure}  
  \centerline{\includegraphics[width=7cm]{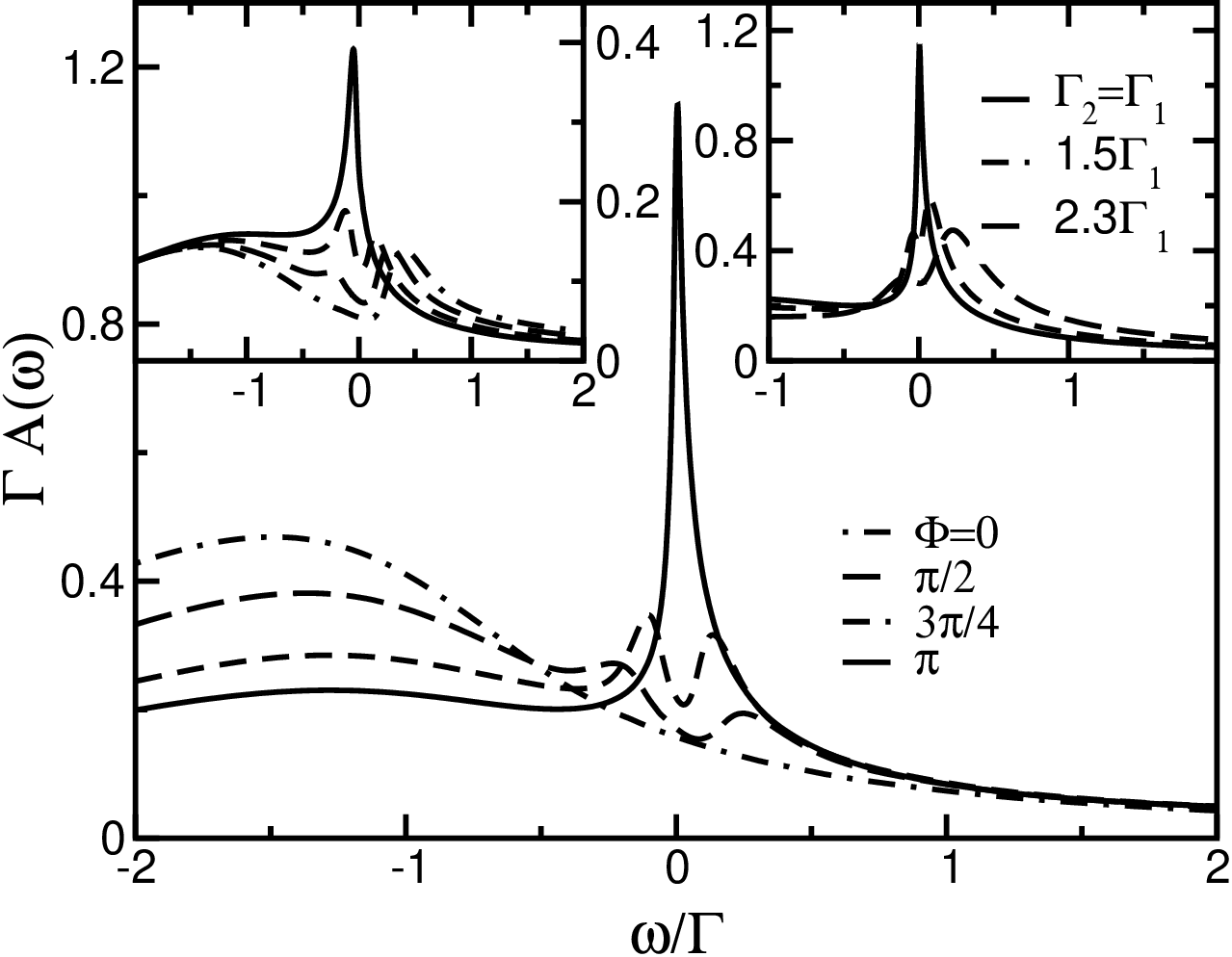}}
        \caption{Effect of a finite AB phase on the single particle spectrum. 
The spectral density of the $f_1$ level is shown for 
$\epsilon_1 = -1.6 \Gamma$, $\delta\epsilon = 0$, $U = 8.1 \Gamma$, 
and $T=0$. 
Left inset: Partial spectral density corresponding 
to the level $c_1$. Same parameters as above, but with a finite 
level splitting $\delta\epsilon = 0.08 \Gamma$.
Right inset: $f_1$ spectral density for different broadening strengths 
$\Gamma_1 \ne \Gamma_2$ of levels $c_1$ and $c_2$.}
        \label{figure3}
\end{figure}
\begin{figure}
  \centerline{\includegraphics[width=7cm]{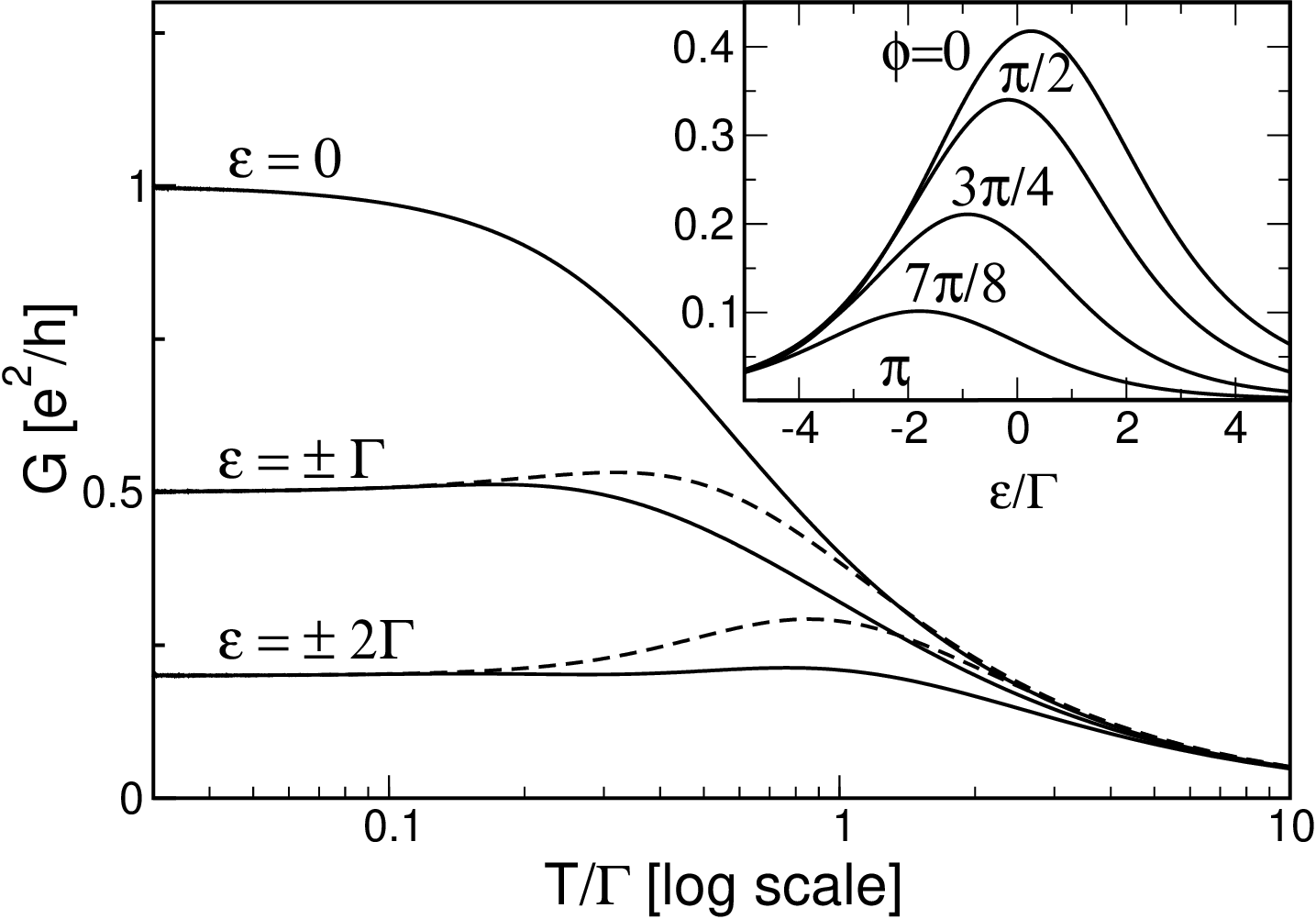}}
        \caption{
          The linear conductance for $\delta
          \epsilon=0$, $D=50 \Gamma$, and $U=\infty$. Main panel:
          $T$-dependence with $\phi=0$ (dashed lines indicate the positive
          energies). Inset: $\phi$-dependence for $T=\Gamma$.
} 
        \label{figure4}
\end{figure}

\begin{references}
\bibitem[*]{new1}{New address: Theoretische Physik A, RWTH Aachen, D-52056
    Aachen, Germany} 

\bibitem{curacao}
{{\em Mesoscopic~Electron~Transport}, edited by L.L.~Sohn, L.P.~Kouwenhoven,
and G.~Sch\"on (Kluwer, Dordrecht,1997); {\em Single Charge Tunneling}, edited
by H.~Grabert and M.H.~Devoret, (Plenum, New York, 1991); D.V.~Averin and
K.K.~Likharev, in {\em Mesoscopic Phenomena in Solids}, ed. B.L.~Altshuler,
P.A.~Lee, and R.A.~Webb (Elsevier, Amsterdam, 1991).}

\bibitem{kondo-exp}
D.~Goldhaber-Gordon {\em et al.}, Nature {\bf 391}, 156 (1998);
S.M.~Cronenwett {\em et al.}, Science {\bf 281}, 540 (1998);
J.~Schmid {\em et al.}, Phys. Rev. Lett. {\bf 84}, 5824 (2000).
W.~G.~van der Wiel {\em et al.}, Science {\bf 289}, 2105 (2000).

\bibitem{simmel}
F.~Simmel {\em et al.}, Phys. Rev. Lett. {\bf 83}, 804 (1999).

\bibitem{kondo-theo}
L.I.~Glazman and M.E.~Raikh, Pis'ma Zh.~ \'Eksp.~Teor.\ {\bf 47}, 378 (1988)
[JETP Lett.\ {\bf 47}, 452 (1988)]; 
T.K.~Ng, P.A.~Lee, Phys. Rev. Lett. {\bf 61}, 1768 (1988); Y.~Meir,
N.S.~Wingreen, and P.A.~Lee, Phys. Rev. Lett. {\bf 70}, 2601 (1993).

\bibitem{yacoby}
A.~Yacoby, M.~Heiblum, D.~Mahalu, and H.~Shtrikman, Phys. Rev. Lett. {\bf 74},
4047 (1995); R.~Schuster {\em et al.}, Nature {\bf 385}, 417 (1997).

\bibitem{loss}
D.~Loss and E.V.~Sukhorukov, Phys. Rev. Lett. {\bf 84}, 1035 (2000).

\bibitem{holleitner-etal}
A.W.~Holleitner {\em et al.}, Phys.\ Rev.\ Lett.\ in press; cond-mat/0011044.

\bibitem{koenig-gefen}
J.~K\"onig, Y. Gefen, and G. Sch\"on, Phys. Rev. Lett. {\bf 81}, 4468 (1998).

\bibitem{pohjola}
T.~Pohjola {\em et al.}, Europhys. Lett. {\bf 40}, 189-194 (1997). 

\bibitem{inoshita}
T.~Inoshita, A. Shimizu, Y. Kuramoto, and H. Sakaki, Phys. Rev. B {\bf 48},
14725 (1993). 

\bibitem{izumida1}
W.~Izumida, O. Sakai, and Y. Shimizu, J. Phys. Soc. Jpn. {\bf 67}, 2444 (1998).

\bibitem{izumida2}
W.~Izumida, O. Sakai, and Y. Shimizu, J. Phys. Soc. Jpn. {\bf 66}, 717 (1997).

\bibitem{schoeller-RG}
H.~Schoeller and J.~K\"onig, Phys. Rev. Lett. {\bf 84}, 3686 (2000);
H.~Schoeller, in {\em Low-Dimensional Systems}, edited by  T.~Brandes,
(Springer, Berlin, 1999), p.137;
H.~Schoeller, Habilitation thesis, Karlsruhe 1997.

\bibitem{NRG}
K.~G.~Wilson, Rev. Mod. Phys. {\bf 47}, 773 (1975); 
T.A.~Costi, A.C. Hewson, and V. Zlati\'{c}, 
J. Phys.:Cond. Mat. {\bf 6}, 2519 (1994); 
W.~Hofstetter, Phys. Rev. Lett. {\bf 85}, 1508 (2000).

\bibitem{hewson}
A.C.~Hewson, {\em The Kondo Problem to Heavy Fermions}, 
(Cambridge University Press, Cambridge, 1993).

\bibitem{matveev}
K.A.~Matveev, Zh.~\'Eksp.\ Teor.\ Fiz.\ {\bf 99} 1598 (1991) [Sov. Phys. JETP
{\bf 72}, 892 (1991)]. 

\bibitem{zawadowski}
K.~Vladar and A.~Zawadowski, Phys. Rev. B {\bf 28}, 1564 (1983);
A.~Zawadowski, J.~v.~Delft, and D.C.~Ralph, Phys. Rev. Lett. {\bf 83}, 2632
(1999).

\bibitem{ruderman}M.A.~Ruderman and C.~Kittel, Phys. Rev. {\bf 96}, 
99 (1954).

\bibitem{exsolv}This constitutes a special case, for which exact results are
  shown. 
\end{references}
\end{document}